\documentclass[aps,prapp,10pt,superscriptaddress,twocolumn,longbibliography]{revtex4-1}

\usepackage{amsmath}
\usepackage{amssymb}
\usepackage{amsfonts}
\usepackage{ascmac}
\usepackage{color}
\usepackage{comment}
\usepackage{here}
\usepackage{physics}
\usepackage{qcircuit}
\usepackage{siunitx}
\usepackage{tikz}
\usepackage{ulem}
\usepackage{tabularx}

\usepackage{hyperref}
\hypersetup{
    colorlinks=true,
    linktocpage=true,
    linkcolor=blue,
    filecolor=magenta,
    urlcolor=cyan,
    bookmarks=true,}
    
\graphicspath{{figures/}}

\usepackage{xcolor}
\usepackage{ulem}
\DeclareRobustCommand{\Erase}{\bgroup\markoverwith{\textcolor{red}{\rule[.5ex]{2pt}{0.4pt}}}\ULon}

\DeclareRobustCommand{\Erasesecond}{\bgroup\markoverwith{\textcolor{green}{\rule[.5ex]{2pt}{0.4pt}}}\ULon}

\begin{document}

\title{Active Initialization Experiment of Superconducting Qubit Using Quantum-circuit Refrigerator}

\author{Teruaki Yoshioka}
\affiliation{Department of Physics, Tokyo University of Science, 1--3 Kagurazaka, Shinjuku, Tokyo 162--0825, Japan}
\affiliation{RIKEN Center for Quantum Computing (RQC), 2--1 Hirosawa, Wako, Saitama 351--0198, Japan}
\affiliation{National Institute of Advanced Industrial Science and Technology (AIST), National Metrology Institute of Japan (NMIJ), 1-1-1 Umezono, Tsukuba, Ibaraki 305-8563, Japan}

\author{Hiroto Mukai}
\affiliation{Department of Physics, Tokyo University of Science, 1--3 Kagurazaka, Shinjuku, Tokyo 162--0825, Japan}
\affiliation{RIKEN Center for Quantum Computing (RQC), 2--1 Hirosawa, Wako, Saitama 351--0198, Japan}

\author{Akiyoshi Tomonaga}
\affiliation{Department of Physics, Tokyo University of Science, 1--3 Kagurazaka, Shinjuku, Tokyo 162--0825, Japan}
\affiliation{RIKEN Center for Quantum Computing (RQC), 2--1 Hirosawa, Wako, Saitama 351--0198, Japan}

\author{Shintaro Takada}
\affiliation{National Institute of Advanced Industrial Science and Technology (AIST), National Metrology Institute of Japan (NMIJ), 1-1-1 Umezono, Tsukuba, Ibaraki 305-8563, Japan}

\author{Yuma Okazaki}
\affiliation{National Institute of Advanced Industrial Science and Technology (AIST), National Metrology Institute of Japan (NMIJ), 1-1-1 Umezono, Tsukuba, Ibaraki 305-8563, Japan}

\author{Nobu-Hisa Kaneko}
\affiliation{National Institute of Advanced Industrial Science and Technology (AIST), National Metrology Institute of Japan (NMIJ), 1-1-1 Umezono, Tsukuba, Ibaraki 305-8563, Japan}

\author{Shuji Nakamura}\email{shuji.nakamura@aist.go.jp}
\affiliation{National Institute of Advanced Industrial Science and Technology (AIST), National Metrology Institute of Japan (NMIJ), 1-1-1 Umezono, Tsukuba, Ibaraki 305-8563, Japan}

\author{Jaw-Shen Tsai}\email{tsai@riken.jp}
\affiliation{Department of Physics, Tokyo University of Science, 1--3 Kagurazaka, Shinjuku, Tokyo 162--0825, Japan}
\affiliation{RIKEN Center for Quantum Computing (RQC), 2--1 Hirosawa, Wako, Saitama 351--0198, Japan}

\begin{abstract}
The initialization of superconducting qubits is one of the essential techniques for the realization of quantum computation. 
In previous research, initialization above 99\% fidelity has been achieved at 280 ns. Here, we demonstrate the rapid initialization of a superconducting qubit with a quantum-circuit refrigerator (QCR). Photon-assisted tunneling of quasiparticles in the QCR can temporally increase the relaxation time of photons inside the resonator and helps release energy from the qubit to the environment. Experiments using this protocol have shown that 99\% of initialization time is reduced to 180 ns. This initialization time depends strongly on the relaxation rate of the resonator, and faster initialization is possible by reducing the resistance of the QCR, which limits the ON/OFF ratio, and by strengthening the coupling between the QCR and the resonator.
\end{abstract}

\maketitle

\section{Introduction}
Quantum computation to solve some computational problems on time scales much faster than those of conventional classical computers has been a major challenge over the past few decades~\cite{nakamura_coherent_1999, Arute2019Quantum, GQA2021Exponential, PhysRevLett.127.180501}. 
To realize a quantum computer, initialization, gate operation, and readout of qubit states are the essential processes~\cite{divincenzo2000physical,Kwon2021Gate}. To date, although relatively high speed and high fidelity have been achieved in the manipulation and readout of qubits, a further improvement in the initialization time is required experimentally. 

In various proposals for the initialization of a superconducting qubit~\cite{PhysRevLett.110.120501,PhysRevApplied.10.044030, reed2010fast, PhysRevB.101.235422, PhysRevApplied.17.044016}, the quantum-circuit refrigerator (QCR)~\cite{tan2017quantum} is a promising candidate for fast initialization~\cite{PhysRevB.101.235422}. 
The fastest method of initializing qubits, proposed in Ref.~\cite{PhysRevB.101.235422}, suggests that $6\,\si{\nano\second}$ initialization can be achieved theoretically. However, in that method, a qubit should be connected directly to a QCR. 
Thus, the coherence of the qubits is markedly decreased.

In this study, to retain the coherence of a qubit, the qubit is coupled to a QCR system through a resonator. 
Theoretically, the initialization system with a resonator connected to a QCR only reduces the qubit coherence time by less than 1\%.  
A QCR has a superconductor--insulator--normal metal--insulator--superconductor (SINIS) junction~\cite{masuda2018observation, pekola2010environment}, which absorbs energy spontaneously from the circuit system via photon-assisted tunneling~\cite{ingold1992charge}.  

Our strategy for a rapid and efficient qubit initialization is based on three previous works~\cite{sevriuk2019fast, PhysRevLett.121.060502, PhysRevB.100.134505}.
Refs.~\cite{sevriuk2019fast, PhysRevB.100.134505} showed that a QCR reduces the photon relaxation time of a superconducting resonator by releasing the energy of a resonator to the environment, and Refs.~\cite{PhysRevApplied.10.044030,PhysRevLett.121.060502} showed that two microwave pulses convert a population of excited states of a transmon qubit into photons inside the resonator, which are eventually released to the environment through cavity photon loss.
Combining these two circuits, i.e., a qubit coupled to a tunable photon-relaxation rate resonator with the QCR, we achieve qubit initialization using the QCR with 99\% accuracy in $180\, \si{\nano\second}$.
This is the first experimental realization of qubit initialization using microwaves in combination with a QCR and a resonator in contrast to Ref.~\cite{sevriuk_initial_2022} where a transmon qubit was reset by a directly coupled QCR.~\cite{sevriuk2019fast, PhysRevB.100.134505,sevriuk_initial_2022}.

\section{General description}
\begin{figure*}[t]
\centering
\includegraphics[width=150mm]{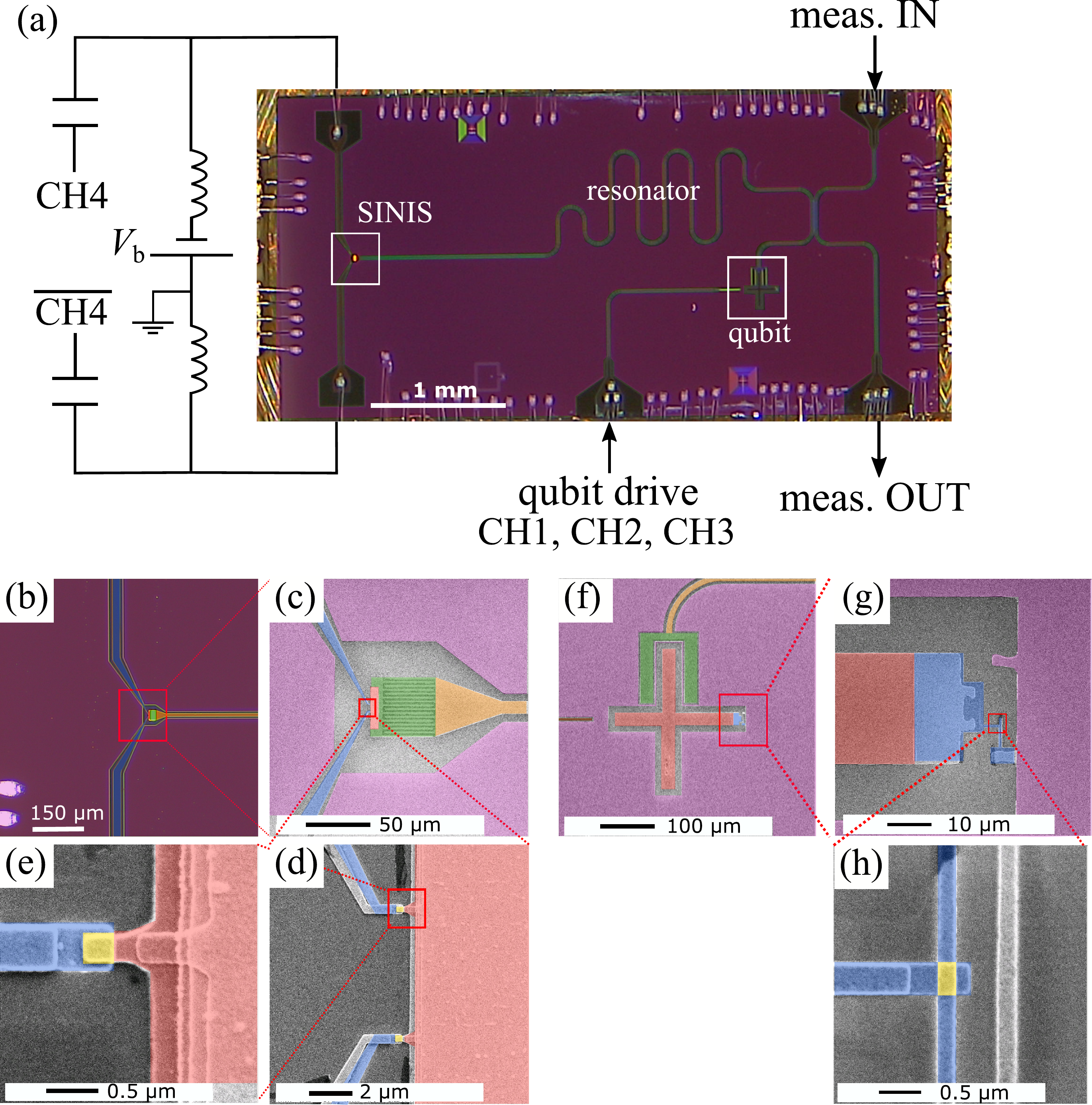}
        \caption{
        (a) Optical image of measured device.
        Conductor of the chip made of Nb (purple) and etched to form the resonator, gaps of the lead line for SINIS bias, interdigital capacitance, and ports. The labels on the wires indicate the parts shown in the figure shown below. $\mathrm{CH_4}$ is the port that outputs a bias pulse to be applied to SINIS. $\overline{\mathrm{CH_4}}$ is the port that outputs the signal of $\mathrm{CH_4}$ with opposite sign. The DC and AC lines of the SINIS bias line are formed by bias-T.
                (b) (c) Optical and SEM images of the region around the QCR. The resonator (orange) is coupled to the SINIS by an interdigital capacitor (green). The SINIS is deposited by the double-angle evaporation of Cu (red) and Al (blue). These devices are surrounded by a Nb ground (purple). 
        (d) SEM image focused around the SINIS junction (red square in (c)). 
        Al (blue) and Cu (red) are used for the superconductor and normal metal of the SINIS, respectively. The insulator is $\mathrm{AlO}_x$ (yellow).
        (e) SEM image of the NIS junction (the same color as that in (d)).
        (f) (g) SEM image of the area around the qubit. The resonator (orange) is coupled to the qubit by a capacitor (green). The Josephson junction is deposited by double-angle evaporation of Al--Al (blue) similar to SINIS junctions. At this time, the natural oxide film of the Nb island (red) is removed by Ar milling to form a superconducting contact between the Nb island and Al. These devices are surrounded by Nb ground (purple).
        (h) SEM image focused around the Josephson junction (red square in (g)). Al--Al (blue) is used as the superconductor of the qubit. The insulator is $\mathrm{AlO}_x$ (yellow).
        }
\label{fig:sample}
\end{figure*}
In this study, the superconducting circuit is composed of a transmon qubit and a coplanar waveguide resonator for readout, as well as an SINIS junction for the QCR. The superconducting resonator is capacitively coupled to the transmon qubit and the SINIS at each end. To readout the state of the qubit, the transmission line is coupled to the resonator (shown in Fig.~\ref{fig:sample}).
The substrate sputtered with 50 nm Nb on a 3 inch Si wafer was cut into 20 mm square pieces and fabricated by a process used for fabricating semiconductors. First, a resonator, capacitances, and a transmission line were drawn on the substrate by photolithography, and Nb was etched by reactive ion etching. We then fabricated a qubit. The process involved drawing the mask by electron beam lithography and using the double-angle shadow evaporation method. The first 40-nm-thick Al layer was deposited, followed by the second 60-nm-thick Al layer. After the first Al layer was deposited, oxygen was allowed to flow into the chamber to naturally form an oxide film of Al. Once the qubit was fabricated, it was diced into 2.5$\times$5.0 mm chips. The SINIS junction is then fabricated similarly to the fabrication of the qubit. The first 40-nm-thick Al layer was deposited, followed by the second 60-nm-thick Cu layer.

The Hamiltonian system of the circuit excluding the SINIS junction is described as
\begin{align} \label{reset_hamiltonian}
\hat{H}/\hbar = 
 & \,\, \omega_\mathrm{r}\hat{a}^\dagger\hat{a}+\omega_\mathrm{ge}\hat{b}^\dagger\hat{b} \nonumber \\
 & + \frac{\alpha}{2}\hat{b}^\dagger\hat{b}^\dagger\hat{b}\hat{b} +\lambda(\hat{b}^\dagger\hat{a}+\hat{b}\hat{a}^\dagger) \nonumber \\
 & + \frac{g_\mathrm{Rabi}}{\sqrt{2}}(
      \hat{b}^\dagger\hat{b}^\dagger\hat{a}{\rm e}^{i\omega_\mathrm{f0g1}t} 
    + \hat{a}^\dagger\hat{b}\hat{b}{\rm e}^{-i\omega_\mathrm{f0g1}t} ) \\
 & + \frac{{\it\Omega}_\mathrm{Rabi}}{\sqrt{2}}(\hat{b}{\rm e}^{i\omega_\mathrm{ef}t} + \hat{b}^\dagger{\rm e}^{-i\omega_\mathrm{ef}t}) \nonumber~,
\end{align}
where $\omega_\mathrm{r}/2\pi$, $\omega_\mathrm{ge}/2\pi$, and $\omega_\mathrm{ef}/2\pi$ represent the resonator frequency and the qubit transition frequencies between states $\ket{g}$ and $\ket{e}$ and states $\ket{e}$ and $\ket{f}$, respectively. 
Here, $\lambda/2\pi$ represents the coupling frequency between the qubit and the resonator, and $\alpha = \omega_\mathrm{ef} - \omega_\mathrm{ge}$ represents the anharmonicity of the transmon qubit. The term $g_\mathrm{Rabi}/2\pi$ is the Rabi frequency between the $\ket{f,0}$ and $\ket{g,1}$ states, and ${\it\Omega}_\mathrm{Rabi}$ is the Rabi frequency between the $\ket{e,0}$ and $\ket{f,0}$ states.

\begin{figure*}[t]
\centering
\includegraphics[width=180mm]{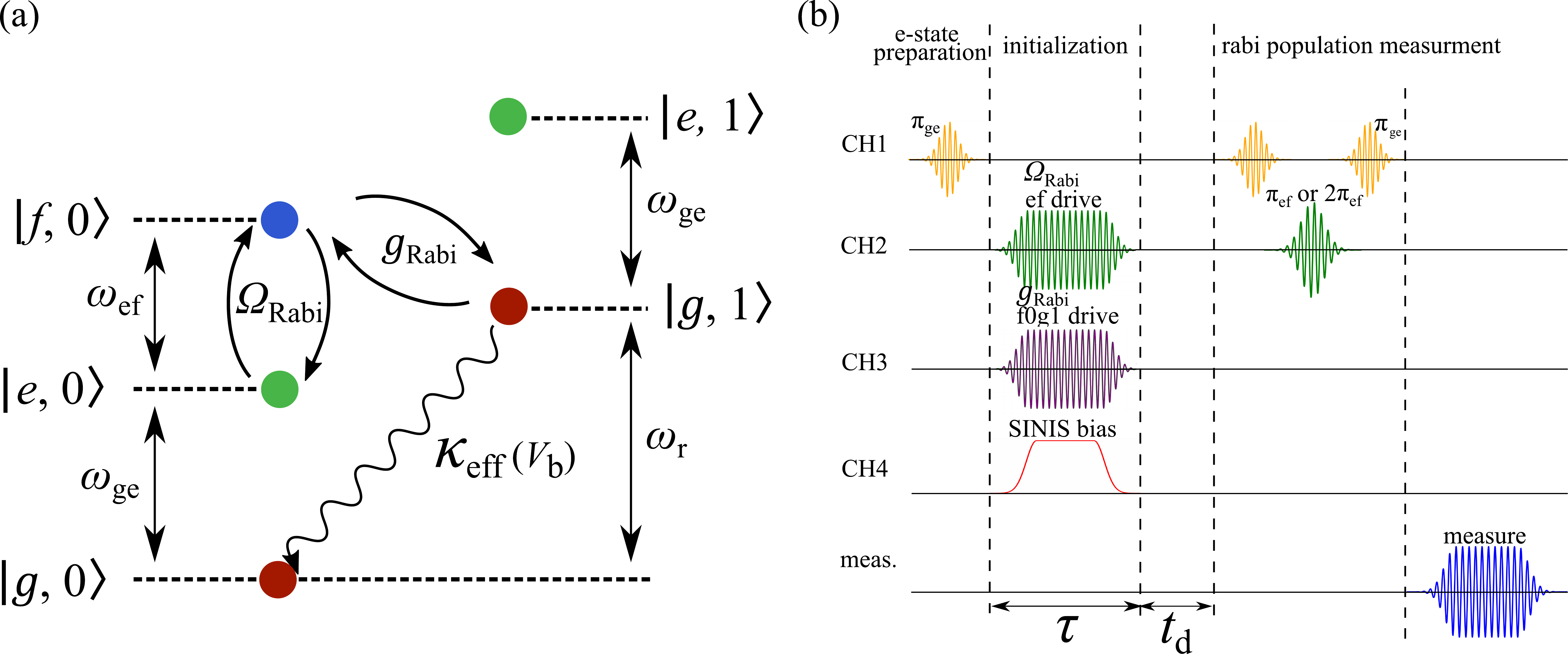}
    \caption{
    (a) Hamiltonian (Eq.~\eqref {reset_hamiltonian}) energy diagram. 
    The Rabi frequencies of the transitions are induced by the drive pulses between $\ket{e,\,0}$ and $\ket{f,\,0}$, $\ket{f,\,0}$ and $\ket{g,\,1}$, respectively.
    (b) Pulse sequence of this experiment from state preparation to initialization and measurement. 
    The lines represent channels of the arbitrary waveform generator (AWG) that generates the waveform. 
    The $\ket{e}$ state is prepared as the pre-initialization state with $\pi_\mathrm{ge}$. 
    Then, an initialization pulse is applied to obtain the occupation of the $\ket{e}$ state in Rabi population measurement (RPM). 
    Here, ${\it\Omega}_\mathrm{Rabi}$ and $g_\mathrm{Rabi}$ are the Rabi frequency between levels shown in (a). A total of four RPMs were carried out, with and without the first $\pi_\mathrm{ge}$ and with CH2 using $\pi_\mathrm{ef}$ or $2\pi_\mathrm{ef}$. 
    }
\label{fig:diagram}
\end{figure*}

The energy diagram of the circuit is shown in Fig.~\ref{fig:diagram}(a) where the eigenstate of the qubit ($n = g,\, e,\, f$) and the Fock state of the resonator ($m=0,\, 1,\, 2,\, \ldots$) are taken into account. The state $\ket{n,\,m}$ represents their product state.
In the energy diagram, to release qubit energy to the environment via a resonator, applying two drive pulses with frequencies $\omega_\mathrm{ef}/2\pi$ (transition frequency between $\ket{e,0}$ and $\ket{f,0}$) and $\omega_\mathrm{f0g1}/2\pi$ (transition frequency between $\ket{f,0}$ and $\ket{g,1}$) enables the unconditional initialization of a qubit based on Ref.~\cite{PhysRevLett.121.060502} 
The qubit state $\ket{e,0}$ will be excited to the $\ket{f,0}$ state by the ${\it\Omega}_\mathrm{Rabi}$ pulse, and the $\ket{f,0}$ energy will be exchanged with that of the $\ket{g,1}$ state by the $g_\mathrm{Rabi}$ pulse; then, the resonator energy will release its photon to the environment at $\kappa_\mathrm{r}$. As a result, the qubit will be initialized by reaching the ground state $\ket{g,0}$.

\begin{table}[b]
  \caption{Sample parameters used in the experiment. Each parameter is defined by premeasurement. The parameters marked with an asterisk are design values.
  }
  \label{table:sample}
  \centering
  \small
 \begin{tabularx}{0.50\textwidth}{Xccc}
    \hline
    Parameter & Symbol & Value\\
    \hline \hline
    Resonator frequency & $\omega_\mathrm{r}/2\pi$  & 6.538 GHz \\
    $g-e$ transition frequency & $\omega_\mathrm{ge}/2\pi$   & 4.663 GHz \\
    $e-f$ transition frequency & $\omega_\mathrm{ef}/2\pi$  & 4.401 GHz \\
    $f0-g1$ transition frequency &  $\omega_\mathrm{f0g1}/2\pi$  & 2.499 GHz \\
    Anharmonicity of qubit & $\alpha/2\pi$  & -261.8 MHz \\
    Resonator--qubit detuning & $\delta_\mathrm{d}/2\pi$  & 1.876 GHz \\
    Qubit lifetime & $T_1$ & 9.6 \si{\micro\second} \\
    Qubit coherence & $T_2^*$ & 2.3 \si{\micro\second} \\
    Josephson energy/Qubit charging energy & $E_\mathrm{J}/E_\mathrm{C}$  & 44.2 \\
    Qubit island capacitance & $C_\mathrm{q}$ & 73.9 fF \\
    Qubit$-$resonator coupling frequency & $\lambda$ & 136 MHz \\
    Bare resonator relaxation rate & $\kappa_\mathrm{r}$  & $2.36\times10^6$ 1/s \\
    SINIS$-$resonator coupling capacitance * & $C_\mathrm{c}$ & 22.7 fF \\
    NIS junction capacitance * & $C_\mathrm{j}$ & 5 fF \\
    SINIS normal-metal-island capacitance * & $C_\mathrm{m}$ & 5 fF \\
    Tunnel resistance of SINIS & $R_\mathrm{T}$  & 72 ${\rm k\Omega}$ \\
    Normal-metal electron temperature & $T_\mathrm{N}$  & $60$ mK \\
    Dynes parameter & $\gamma_\mathrm{D}$   &$1.3\times10^{-4}$  \\
    Energy gap parameter of Al leads & $\Delta$ & 193 $\mathrm{\mu eV}$  \\
    \hline
  \end{tabularx}
\end{table}

For Eq.~\eqref{reset_hamiltonian}, the term $\kappa_\mathrm{r}$ is included as a damping term in the Lindblad equation.
The optimal parameters ${\it\Omega}_\mathrm{Rabi}$ and $g_\mathrm{Rabi}$ to minimize the initialization time (maximize $\kappa_\mathrm{r}$) are obtained to satisfy the relation~\cite{PhysRevLett.121.060502}
\begin{equation}
\label{fast reset condition}
{\it\Omega}_\mathrm{Rabi}=\frac{1}{6}\sqrt{18g^2_\mathrm{Rabi}-\kappa^2_\mathrm{r}}~~\mathrm{and}~~g_\mathrm{Rabi}\geq\sqrt{\frac{2}{27}}\kappa_{\mathrm{r}}.
\end{equation}
In practice, we tune the drive power of two pulses to match the condition.
When the system evolves under this condition, the exponential decay component of initialization rate is maximized and becomes $\kappa_\mathrm{r}/3$.

In this reset scheme, the relaxation rate of the resonator is the main factor that limits the maximum speed of qubit initialization.
However, in a conventional superconducting circuit, the relaxation rate and the coupling $\lambda$ to the qubit are constrained by the design and fabrication, and it will be very difficult to adjust values afterward. 
A high resonator relaxation rate that can be designed for rapid initialization shortens the lifetime of the qubit. 
Thus, there is a trade-off between qubit lifetime and initialization time.

Therefore, to resolve these conflicting objectives, we introduce QCR coupling to the resonator. 
The QCR with the SINIS junction works as a knob to tune the relaxation rate of the resonator, and the SINIS junction can be controlled by voltage $V_\mathrm{b}$~\cite{doi:10.1063/5.0057894}.
When an appropriate bias is applied to the SINIS junction, photon-assisted electron tunneling occurs, causing the SINIS junction to absorb energy from the resonator, thereby enhancing the relaxation rate of the resonator.
The resonator relaxation rate $\kappa_\mathrm{r}$ becomes 
\begin{equation} \label{eq:kappa}
    \kappa_\mathrm{eff}(V_\mathrm{b}) = \kappa_\mathrm{r} + \delta\gamma_\mathrm{QCR}(V_\mathrm{b}) \,,
\end{equation}
where $\delta\gamma_\mathrm{QCR}$ is an additional relaxation rate enhanced by QCR cooling.
In our pulse schedule shown in Fig.~\ref{fig:diagram}(b), although the qubit coherence is maintained during the control of a qubit (except initialization) since the QCR is OFF (bias is not activated), the relaxation rate of the resonator is enhanced when the QCR is ON (bias activated) during the initialization process together with qubit initialization drives.

\section{Quantum-circuit refrigerator}
\begin{figure*}[t]
\centering
\includegraphics[width=180mm]{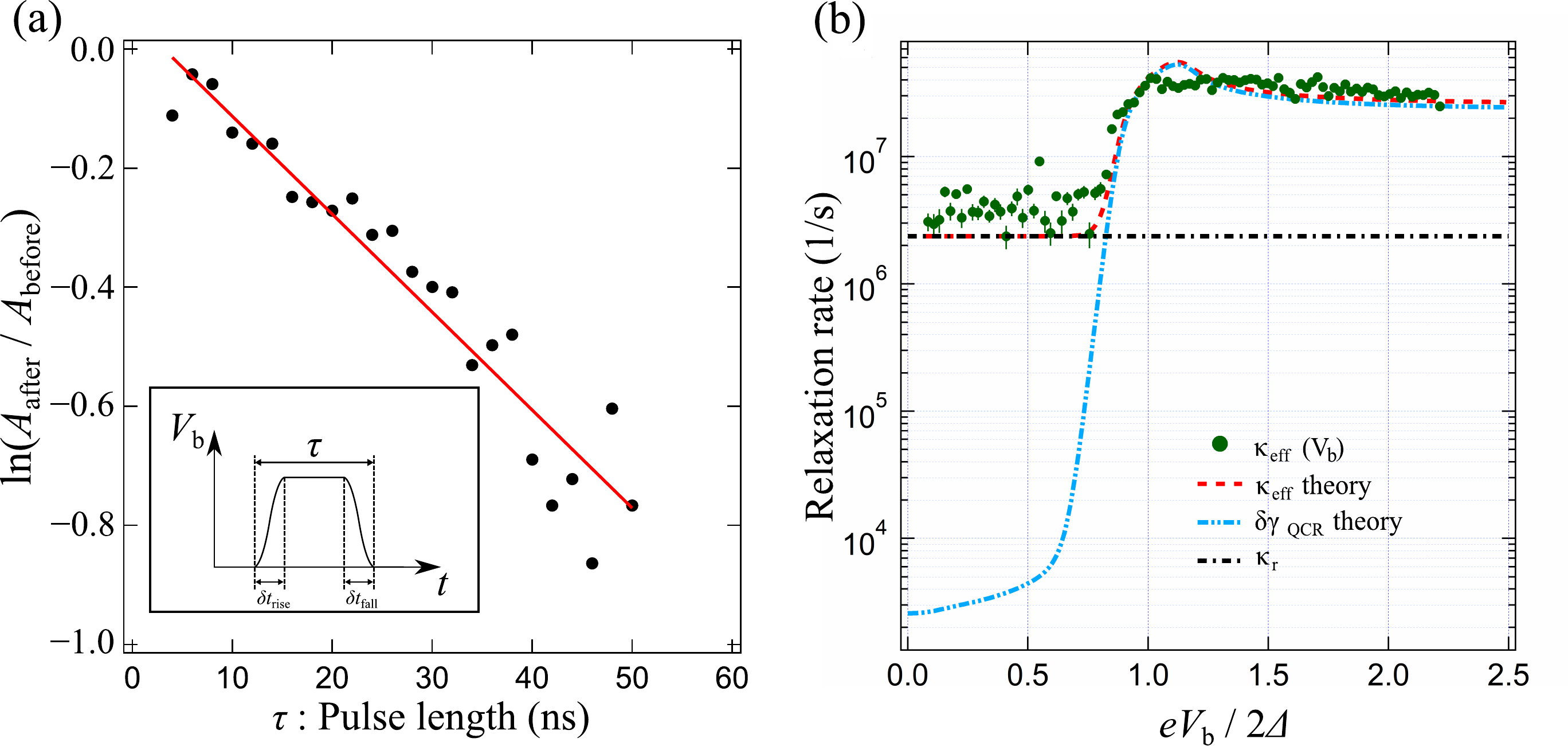}
    \caption{
        (a) 
        Experimental result of Eq.~\eqref{kappa} for the pulse length of the bias voltage applied to the QCR. (black dots) The slope of the function that linearly fits this plot from Eq.~\eqref{kappa} corresponds to $\delta\gamma_{\rm QCR}$. The bias voltage applied to the QCR when this experiment was performed was $eV_\mathrm{b}/2{\it\Delta} = 1.03$. The inset shows the shape of the QCR pulse with pulse length $\tau$. The rise and fall time of the pulse is $\delta t_\mathrm{rise} = \delta t_\mathrm{fall} = 2.5$ ns in the shape of a Gaussian function.
        (b) Relaxation rate of the resonator as a function of bias voltage for SINIS.
        The graph shows the measurement and calculation results of the relaxation rate of the resonator ${\delta\gamma}_\mathrm{QCR}$ dependent on the bias voltage $V_b$ normalized by $e/2{\it\Delta}$.
        The measured relaxation rate of the resonator when the QCR is ON is shown as a green dots and that when the QCR is OFF is shown as a black dot--dash line. 
        The calculated relaxation rate of the resonator as a function of bias voltage for the QCR using Eq.~\eqref{resonator_transition} is shown as a blue dot--dot--dash line.
        The theoretical effective resonator relaxation rate $\kappa_\mathrm{eff~theory}=\kappa_\mathrm{r}+\delta\gamma_\mathrm{QCR~theory}$ is shown as a red dashed line.
        }
\label{fig:kappa}
\end{figure*}

We measure the effective resonator relaxation rate $\kappa_\mathrm{eff}$ depending on the SINIS bias voltage $V_\mathrm{b}$, and the measurement results are plotted in Fig.~\ref{fig:kappa}(a). Below the specific bias point, the relaxation rate is almost constant with background noise. Beyond the bias point, the relaxation rate increases almost by one order of magnitude. Thus, the resonator relaxation is controllable by SINIS bias.

To characterize the relaxation rate of the resonator for $\kappa_\mathrm{eff}$ when the QCR is ON and OFF, two measurements are performed and the results are shown in Fig.~\ref{fig:kappa}. After the readout pulse of frequency $\omega_\mathrm{r}$ is turned OFF, a square pulse of $V_\mathrm{b}$ with the duration $\tau$ is applied to the SINIS while the resonator is naturally relaxed.
When the QCR is OFF (shown in Fig.~\ref{fig:kappa}(a)), that is, SINIS is not biased or biased to a sufficiently small degree to prevent electrons from photon-assisted tunneling, the natural decay of the resonator is obtained as $\kappa_{\rm r} = 2.36\mathrm{\times10^6~1/s}$ by fitting the signal from the resonator at a low probe power~\cite{sevriuk2019fast}. 

On the other hand, when the QCR is ON, that is, SINIS is further biased than the threshold of the tunneling, the relaxation rate can be calculated from the ratio of the signal amplitude of the resonator before applying a QCR bias pulse ($A_\mathrm{before}$) to that after applying ($A_\mathrm{after}$)  as~\cite{sevriuk2019fast}
\begin{equation}
\begin{split}
\label{kappa}
\frac{A_\mathrm{after}}{A_\mathrm{before}} = &\,\, {\rm exp} \Big\{-\frac{1}{2}[\delta\gamma_{\rm QCR}(\tau-\delta t_{\rm rise}-\delta t_{\rm fall})\\
& +\delta\gamma_{\rm QCR}^{\rm rise/fall}(\delta t_{\rm rise}+\delta t_{\rm fall})+\kappa_{\rm r}\delta t_\mathrm{ab}]\Big\}~,
\end{split}
\end{equation}
where $\delta t_\mathrm{ab}$ is the measurement delay time between $A_\mathrm{before}$ and $A_\mathrm{after}$. 
The applied bias pulse is formed as a flat top Gaussian, which has the pulse length $\tau$, and the times of pulse rise $\delta t_{\rm rise}$ and pulse fall $~\delta t_{\rm fall}$. Both are fixed at 2.5 ns in this experiment.

At each bias voltage $V_\mathrm{b}, \delta\gamma_{\rm QCR}$ is measured as the amount of change in the relaxation rate of the resonator before and after the QCR ON pulse (shown in Fig.~\ref{fig:kappa}(b)).
$\delta\gamma_{\rm QCR}^{\rm rise/fall}$ is also measured as the amount of change in the relaxation rate of the resonator during the QCR ON pulse rises and falls. 
We sweep the pulse length $\tau$ and measure $A_{\rm after}/A_{\rm before}$. 
The slope of ${\rm ln}(A_{\rm after}/A_{\rm before})$ depending on $\tau$ is equal to $\delta\gamma_{\rm QCR}$ shown in Fig.~\ref{fig:kappa}(a).
In addition, the fit intercept represents $\delta\gamma_{\rm QCR}^{\rm rise/fall}$.
Then, by measuring the rate in the $eV_\mathrm{b}/2{\it\Delta}$ range from 0 to 2.2, we obtain the data in Fig.~\ref{fig:kappa}(a) for $\kappa_\mathrm{eff}$ by calculating Eq.~\eqref{eq:kappa}.  

Next, we estimate $\delta\gamma_\mathrm{QCR}$ theoretically.
The transition rate between the $m$ and $m'$ Fock states of the resonator coupled to SINIS can be written as
\begin{align} \label{resonator_transition}
    {\it\Gamma}_\mathrm{m,m'}(V_\mathrm{b}) = M^2_\mathrm{m,m'}\frac{2R_\mathrm{K}}{R_\mathrm{T}}
    [& \mathcal{F}(eV_\mathrm{b} + \hbar\omega_\mathrm{r}\ell-E_N) \\
     & + \mathcal{F}(-eV_\mathrm{b} + \hbar\omega_\mathrm{r}\ell-E_N)]\,, \notag 
\end{align}

where $R_\mathrm{K}=h/e^2$ is the resistance quantum, $R_\mathrm{T}$ is the tunnel resistance obtained by ${\it I}$--${\it V}$ measurement of SINIS, $M_\mathrm{m, m'} $ is the element of the transition matrix, $\ell = m-m'$, $E_\mathrm{N}$ is the normal metal island charging energy, and $\mathcal{F}$ is the function on the state distribution~\cite{PhysRevB.101.235422, doi:10.1063/5.0057894, article}.
Using this transition rate (Eq.~\eqref{resonator_transition}) for Fock states $\ket{0}$ and $\ket{1}$, we derive the QCR cooling rate $\delta\gamma_\mathrm{QCR}(V_\mathrm{b})$ as
\begin{equation} \label{eq:gamma_qcr_th}
    \delta\gamma_\mathrm{QCR}(V_\mathrm{b}) = 
    {\it\Gamma}_{0,1}(V_\mathrm{b}) - {\it\Gamma}_{1,0}(V_\mathrm{b}) \,.
\end{equation}
Fig.~\ref{fig:kappa}(b) shows the theoretical prediction of Eq.~\eqref{eq:gamma_qcr_th} using our experimental parameters (shown in Table.~\ref{table:sample}), and $\kappa_\mathrm{eff}$ theoretically estimated using this theoretical $\delta\gamma_\mathrm{QCR}$ reproduces our experimental results well.

For the use of the QCR for initialization, the maximum ON/OFF ratio of $\kappa_\mathrm{eff}$ is required, and it is obtained around $eV_\mathrm{b}/2{\it\Delta} = 1.03$. 
However, in the measured device, the optimal operating bias range is broadened by the pulse shape; therefore, we need to find the best performance point at around this value.
Then, we fine tune the bias and finally choose $eV_\mathrm{b}/2{\it\Delta} = 1.03$ as the optimal bias for SINIS (QCR ON).

\begin{figure}[t]
\centering
\includegraphics[width=85mm]{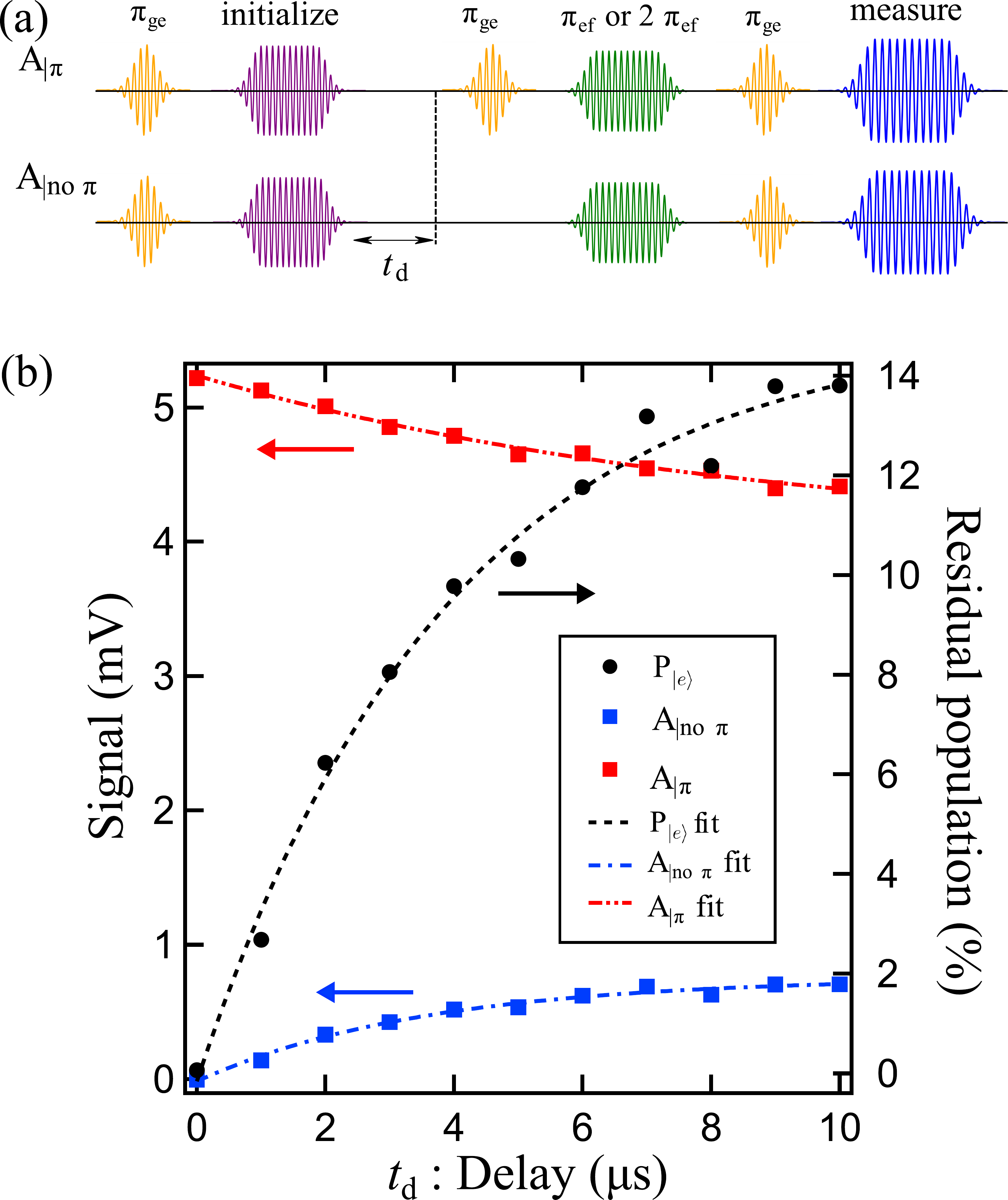}
    \caption{
    (a) RPM measurement scheme. When measuring $A_{|\pi}$, the $\ket{g}$ state is read out by applying $\pi_\mathrm{ge}$ after initialization and then $\pi_\mathrm{ef}$ or $2\pi_\mathrm{ef}$. The amplitude corresponding to the occupation of the $\ket{g}$ state is obtained by this measurement. When measuring $A_{|{\rm no}\,\pi}$, after $\pi_\mathrm{ef}$ or $2\pi_\mathrm{ef}$ is applied after initialization, the $\ket{g}$ state is read out. The amplitude corresponding to the occupation of the $\ket{e}$ state is obtained by this measurement.
    (b) RPM measurement results. $\pi_\mathrm{ge}$ and $\pi_\mathrm{ef}$ have the same amplitude of $44.8~{\rm \mu V}$ and pulse lengths of 102 ns and 90 ns, respectively. $2\pi_\mathrm{ef}$ simply doubled the pulse length. Amplitude values were calculated from AWG output values ($V_\mathrm{pp}$) with the addition of the attenuator and cable attenuation. The time constant to reach thermal equilibrium was determined to be $3.9~{\rm \mu s}$ by sweeping the time from initialization to RPM. $A_{|\pi}$ and $A_{|{\rm no}\,\pi}$ are the measurement results of the scheme in (a). (Left axis) $\mathcal{P}_{\ket{e}}$ is the residual population calculated using Eq.~\eqref{RPM}. (right axis)
    }
\label{fig:RPM}
\end{figure}

\section{Rabi population measurement}
Rabi population measurements~\cite{PhysRevLett.114.240501} are carried out to estimate the population in the ground state after initialization.
By applying a dispersive readout, we calculate the population in the ground state is calculated from the Rabi amplitude.
When we assume that there is no leakage to the $\ket{f}$ state, the population $\mathcal{P}_{\ket{e}}$ is extracted from the measured amplitude of Rabi oscillation between the $\ket{e}$ and $\ket{f}$ states by applying the $\pi$-pulse to excite the $\ket{g}$ state to the $\ket{e}$ state, $A_{|\pi}$, and without applying the $\pi$-pulse $A_{|\text{no}\,\pi}$:

\begin{equation}
\label{RPM}
\mathcal{P}_{\ket{e}} = \frac{A_{|{\rm no}\,\pi}}{A_{|{\rm no}\,\pi}+A_{|\pi}} \,.
\end{equation}

The measurement results are shown in Fig.~\ref{fig:RPM}. The populations of the $\ket{g}$ and $\ket{e}$ states are 85\% and 15\%, respectively. The temperature of the mixing chamber of the dilution refrigerator was 12 mK. Since the dilution refrigerator in this experiment has no shield in the mixing chamber, it is considered that the qubit is thermally excited by radiation from the still.

When the waiting time $t_\mathrm{d}$ from initialization to RPM measurement is swept, an exponential increase in the population can be observed (Fig.~\ref{fig:RPM}(b)).
This represents the thermal excitation time required for the qubit to change from a cooled state to a thermal equilibrium state, which is $3.9~{\rm \mu s}$.

\begin{figure*}[t]
\centering
\includegraphics[width=180mm]{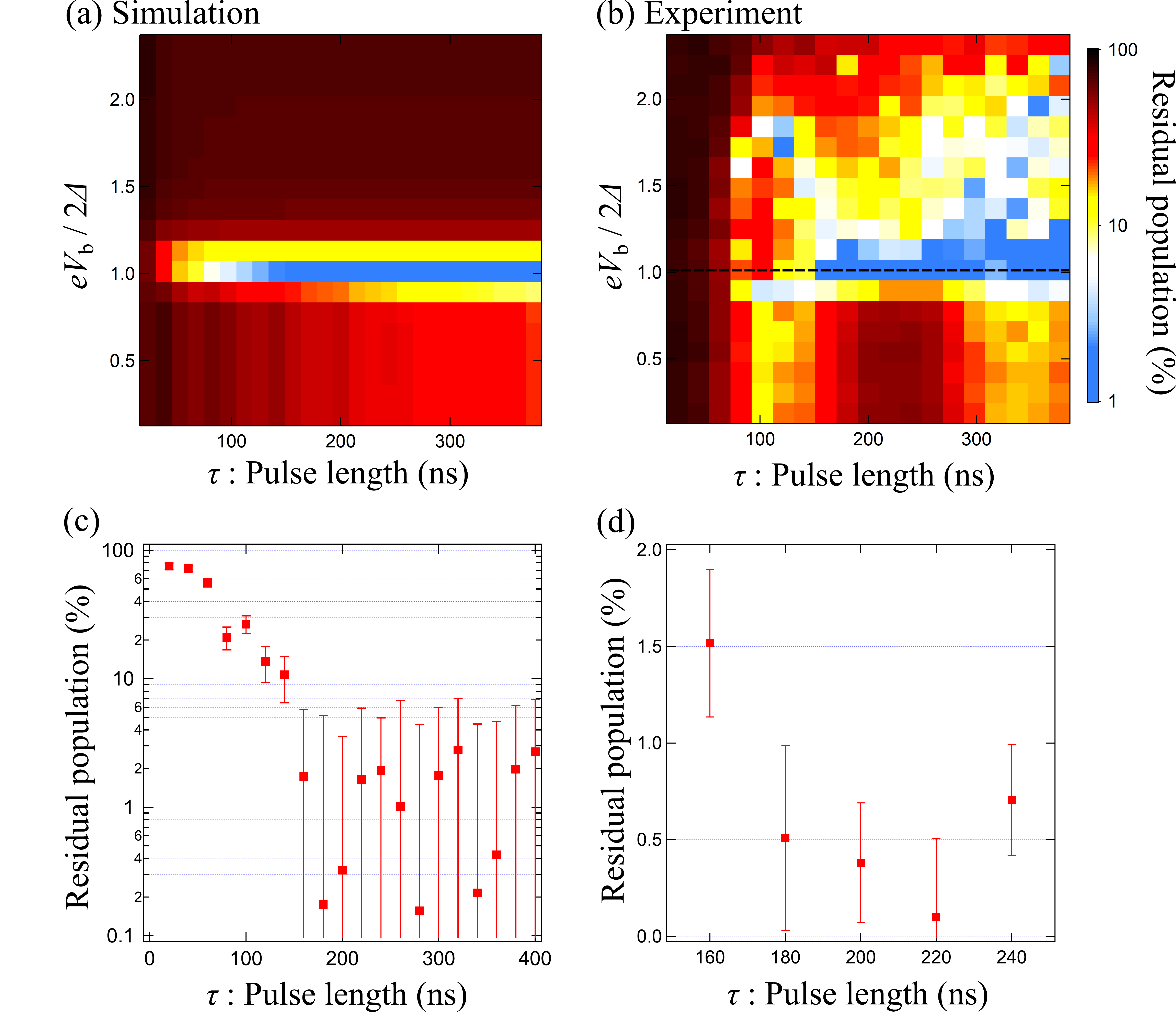}
        \caption{
        (a) Simulation results showing dependence of initialization time on bias voltage of QCR.
        Using data in Table~\ref{table:sample}, we computed the initialization time as described in Ref.~\cite{doi:10.1063/5.0057894}.
        (b) Measurement result showing dependence of initialization time on bias voltage of QCR.
        In the experiment, $g_\mathrm{Rabi}/2\pi$ was fixed at 28.4MHz and the Rabi frequency of the e-f state was set at the value satisfying Eq.~\eqref{fast reset condition}.
        (c) Line plot of the black dashed line in (b) when optimal parameters were used.
        This data is the result of one measurement of 40,000 integral measurements. The error bars in the range of pulse lengths 160--240 ns are $\sqrt{110}$ times the error bars determined in (d). For all other plots, the average of the error bars in (d) multiplied by $\sqrt{110}$ was used. We assumed that this error was limited by the noise of the amplifier in the measurement line and that all plots should show a similar trend.
        (d) Results of 110 repeated measurements of 40,000 integral measurements at around 200 ns.
        }
        \label{fig:result}
\end{figure*}

\section{Initialization}
The scheme of initialization proposed in a previous study \cite{PhysRevLett.121.060502} does not have a structure that dynamically changes the relaxation rate of the resonator. If you increase the relaxation rate too much in an attempt to initialize rapidly, there will be adverse effects such as the inability to read out or the shortening of the qubit lifetime. Therefore, if the QCR is biased at the same time as the two initializing drive pulses are applied, the relaxation rate of the resonator can be increased only at the time of initialization, thus protecting the qubit from the QCR while keeping the relaxation rate of the resonator low except for initializing.

Three types of pulse are used for initialization using the QCR, as shown in Fig.~\ref{fig:diagram}. The ef-drive and f0g1-drive are flat top Gaussian modulated at their corresponding transition frequencies. The QCR is biased by flat top Gaussian pulses from both sides of the source and drain contacts of the QCR. The QCR pulse is generated by AWG at 65 GSa/s, and the rise and fall settings of the pulse are both $\delta t_\mathrm{rise} = \delta t_\mathrm{fall} = 2.5$ ns. The microwave components of the applied line of the QCR pulse supports up to 20 GHz except for the amplifier, and it is considered that the waveform is hardly deformed by the microwave components. As for the voltage amplifier, there is no problem in terms of the pulse shape because the rise and fall time is 1.8 ns, as indicated in the catalog specifications, which is below the set value. It also calibrates the signal output from the AWG to be flat at the output port of the amplifier. Additionally, if the bias applied to the QCR is small due to the rise and fall time effect, the initialization may be faster than that described later because the initialization time in experiments are overestimated.  In the initialization scheme shown in Fig.~\ref{fig:diagram}(b) using microwaves, faster initialization is achieved by maximizing the variable $\kappa_\mathrm{eff}$ shown in Fig.~\ref{fig:kappa} for the QCR.

Currently, the f0g1-drive applies a high-power pulse, so the frequency of a qubit is shifted by an AC Stark shift \cite{PhysRevLett.121.060502}. The QCR bias also causes a Lamb shift at the resonant frequency of the resonator\cite{silveri2019broadband, PhysRevResearch.3.033126}, so both effects must be calibrated.

Figs.~\ref{fig:result}(a) and (b) show the residual population obtained from numerical simulation (a) and experimental (b) results, where $g_\mathrm{Rabi}$ is fixed.
The simulation results are similar to the experimental results.
Fig.~\ref{fig:result}(c) shows the plot at $eV/2{\it\Delta}=1.03$, $g_\mathrm{Rabi}/2\pi=28.4~\mathrm{MHz}$. The excited population decreases as the initialization pulse duration becomes longer, and it reaches almost zero around 200 ns. To estimate the excited population more precisely, we repeated the measurement around 200 ns within the error bar. Fig.~\ref{fig:result}(d) shows the result. The residual population becomes below 99\% at around 180 ns.

\section{Discussion}
The accuracy of the initialization depends on the thermal excitation of qubits and the electron temperature of normal metals.
By using the parameters of this experimental system, one can simply estimate the occupation of $\ket{g}$ after the initialization. Let the transition rate from $\ket{g}$ to $\ket{e}$ be ${\it\Gamma}_{\uparrow}$ and from $\ket{e}$ to $\ket{g}$ be ${\it\Gamma}_{\downarrow}$. The relaxation rate of a qubit can be expressed as $1/T_1={\it\Gamma}_{\uparrow}+{\it\Gamma}_{\downarrow}=$~104$\mathrm{\times10^3~1/s}$. Moreover, the occupation of the excited state in the thermal equilibrium state can be expressed as $\mathcal{P}_{\ket{e}}={\it\Gamma}_{\uparrow}/({\it\Gamma}_{\uparrow}+{\it\Gamma}_{\downarrow})=$~0.15. Since the maximum initialization rate in this experiment is $\kappa_\mathrm{eff}/3=13.8{\rm \times10^6~1/s}$, adding this value to ${\it\Gamma}_{\downarrow}$ to calculate the occupation of $\ket{g}$ yields 99.89\%. On the other hand, the result of this experiment is $99.5\pm{0.5}\%$, which is very close to the calculation result. It is conceivable that an initialization of one order of magnitude lower, 99.9\%, can be achieved by improving the readout measurement precision and reducing the population of the thermally excited states of qubits. 

The resistance of SINIS is the main reason for the limited initialization speed in this experiment. The initialization speed is strongly dependent on the relaxation rate of the resonator, and faster a initialization is possible by increasing the relaxation rate and, accordingly, by increasing $g_\mathrm{Rabi}$. In addition, since the adjustment value of the relaxation rate of the resonator depends inversely on the tunnel resistance of SINIS, it is possible to make it smaller, which enables further high-speed initialization.
It is possible to initialize at an initialization speed of 100 ns or less by optimizing the circuit parameters, such as reducing the resistance of SINIS and increasing the coupling capacitance between SINIS and the resonator.~\cite{doi:10.1063/5.0057894}.

It has also been theoretically shown that directly coupling SINIS to a qubit allows the qubit to be initialized even faster. However, that has the disadvantage of shortening the qubit lifetime by about 12.6\%. If a qubit with a sufficiently long lifetime to tolerate this demerit can be made, the direct-coupling method will also gain advantages.~\cite{sevriuk_initial_2022}

In this experiment, we utilize a transmon robust to charge noise; it has coherence times $T_1 = 9.6 ~\si{\micro\second}$ and $T_2 = 2.3 ~\si{\micro\second}$. With the natural relaxation of this qubit, the initialization of 99\% is calculated as $42 ~\si{\micro\second}$. 
In addition, the $T_1$ used in this study is shorter than those recently reported for superconducting qubits, but a long lifetime qubit can be realized by adding a purcell filter\cite{jeffrey2014fast, reed2010fast} and improving fabrication process for a superconducting qubit. It seems necessary to further investigate whether the scheme using the QCR is practically usable for the initialization by examining the influence of the qubit on QCR in each on/off state. However, we suspect that in the future, when a circuit in which QCRs and qubits coexist is designed with sufficient detuning and fabrication will be improved, T1 of a qubit will be improved, and the negative effect of QCR should be sufficiently reduced.

\section{Conclusion}
We have shown experimentally that the QCR is a powerful tool for accelerating the initialization of superconducting qubits. The initialization time under the optimal condition is 180 ns with $99.5 \pm 0.5\%$.  This technique will help in the realization of quantum computation with quantum error correction, which needs repeated initializations of ancilla qubits. It is also expected to be effective not only in quantum computing but also in research-level experiments. It is expected that repeated experiments can be carried out more efficiently without waiting for the completion of initialization by damping at $T_1$ in experiments where many integrations are carried out.

\section*{Acknowledgment}
This work is based on results obtained from projects JPNP16007 and JPNP20004 subsidized by the New Energy and Industrial Technology Development Organization (NEDO), Japan. This work was also supported by JST CREST (Grant No. JPMJCR1676), Moonshot R \& D (Grant No. JPMJMS2067), and JSPS KAKENHI (Grant Nos. JP18H05258, 20KK0335, and 20H02561).

\section*{Appendix}

\subsection{SINIS ${\it I}$--${\it V}$ measurement}
\begin{figure}[]
\centering
\includegraphics[width=80mm]{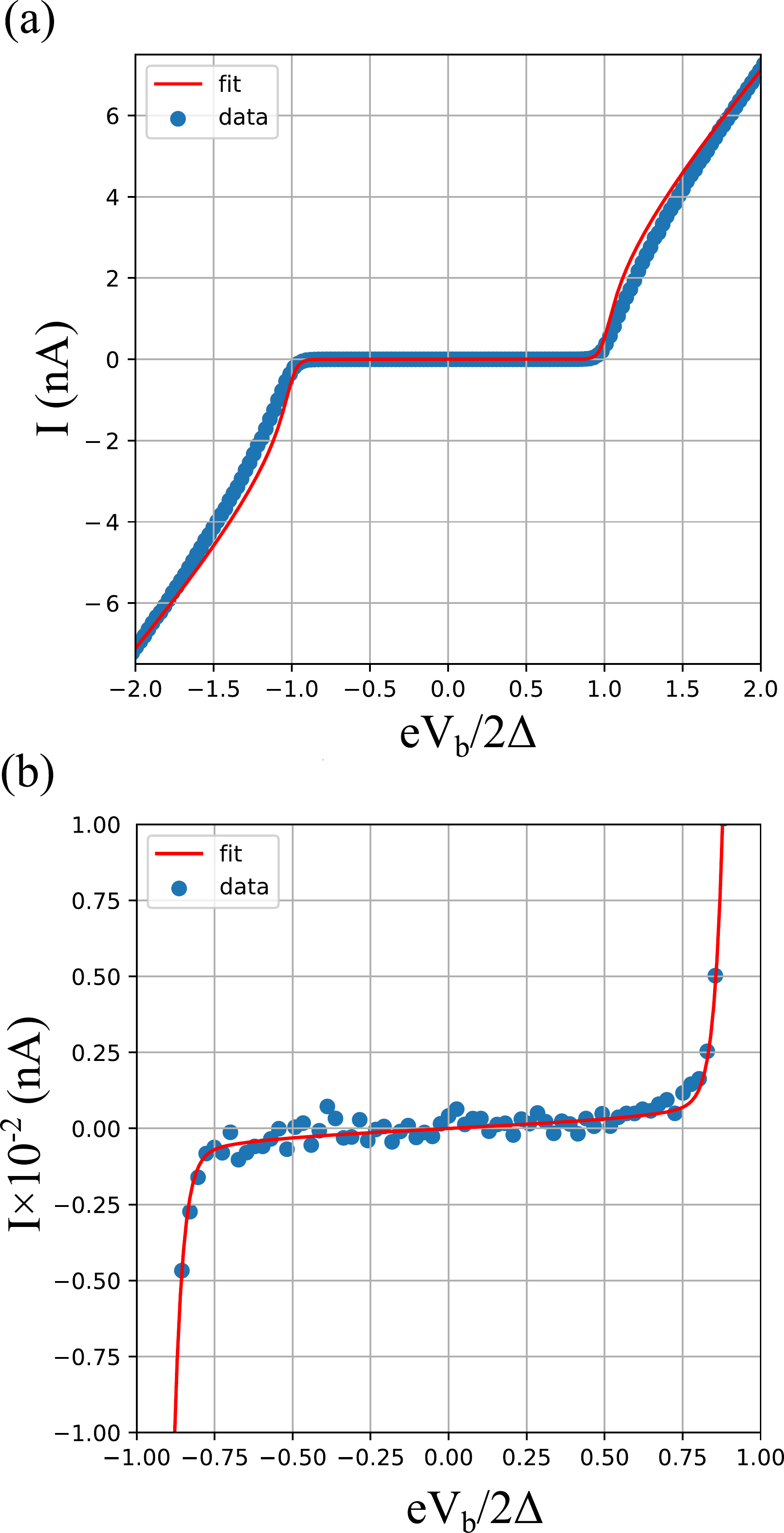}
        \caption{(a) Results of ${\it I}$--${\it V}$ measurement of SINIS using the measurement circuit shown in Fig.~\ref{fig:setup}(b).
        (b) Enlarged view of the plateau region attributable to the superconducting gap in (a).
        }
        \label{fig:IV}
\end{figure}

${\it I}$--${\it V}$ measurements of SINIS are described. To suppress the noise of the voltage source, voltage is applied to the leads at both ends of the SINIS at 1/1000 by a voltage divider. The current flowing through the SINIS junction induced by the bias was amplified using a current amplifier and measured with a digital multimeter. The measurement results are shown in Fig.~\ref{fig:IV}. In addition, 
\begin{equation}
\label{SINIS_I}
I=\frac{1}{2eR_\mathrm{T}}\int dEn_\mathrm{s}(E)[f_\mathrm{N}(E-eV)-f_\mathrm{N}(E+eV)],
\end{equation}
was used for the fitting of Fig.~\ref{fig:IV} \cite{muhonen2012micrometre}. The functions $n_\mathrm{s}$ and $f_\mathrm{N}$ are the quasiparticle density of states of the superconductor and the density of states Fermi function of the normal metal, respectively.

\subsection{Qubit $T_1$ measurement}
\begin{figure}[t!]
\centering
\includegraphics[width=80mm]{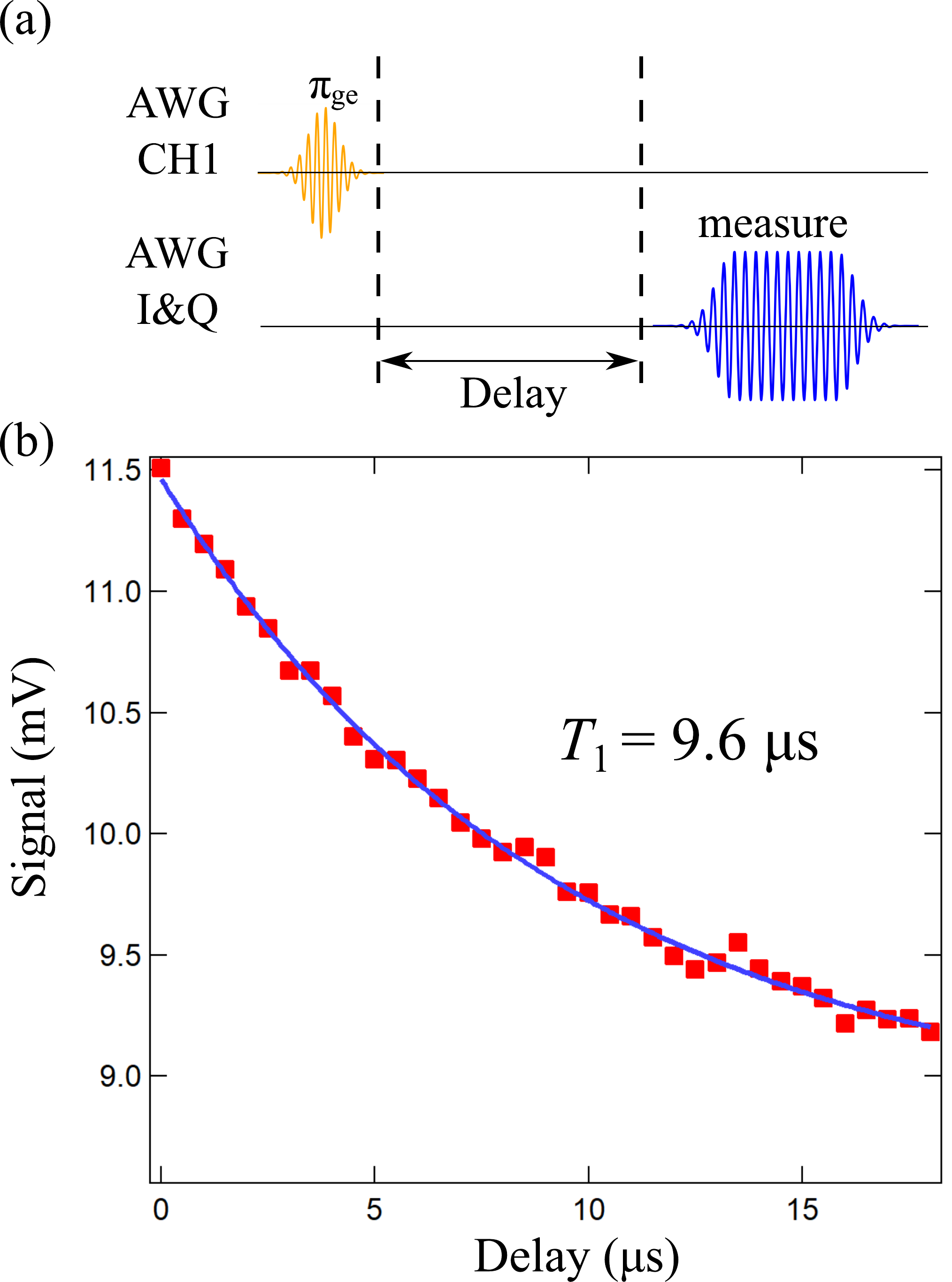}
        \caption{(a) Measurement scheme for the lifetime $T_1$ of a qubit.
        (b) Measurement results of scheme in (a).
        }
        \label{fig:T1}
\end{figure}

We describe lifetime $T_1$ measurements of qubits. The excited state is first generated by a $\pi_\mathrm{ge}$ pulse, as shown in Fig.~\ref{fig:T1}(a). Then, the lifetime $T_1$ of the qubit is measured by varying the delay time. The measurement results are shown Fig.~\ref{fig:T1}(b). This measurement includes the thermal excitation rate.

\subsection{Rabi population measurement}
\begin{figure}[t!]
\centering
\includegraphics[width=78mm]{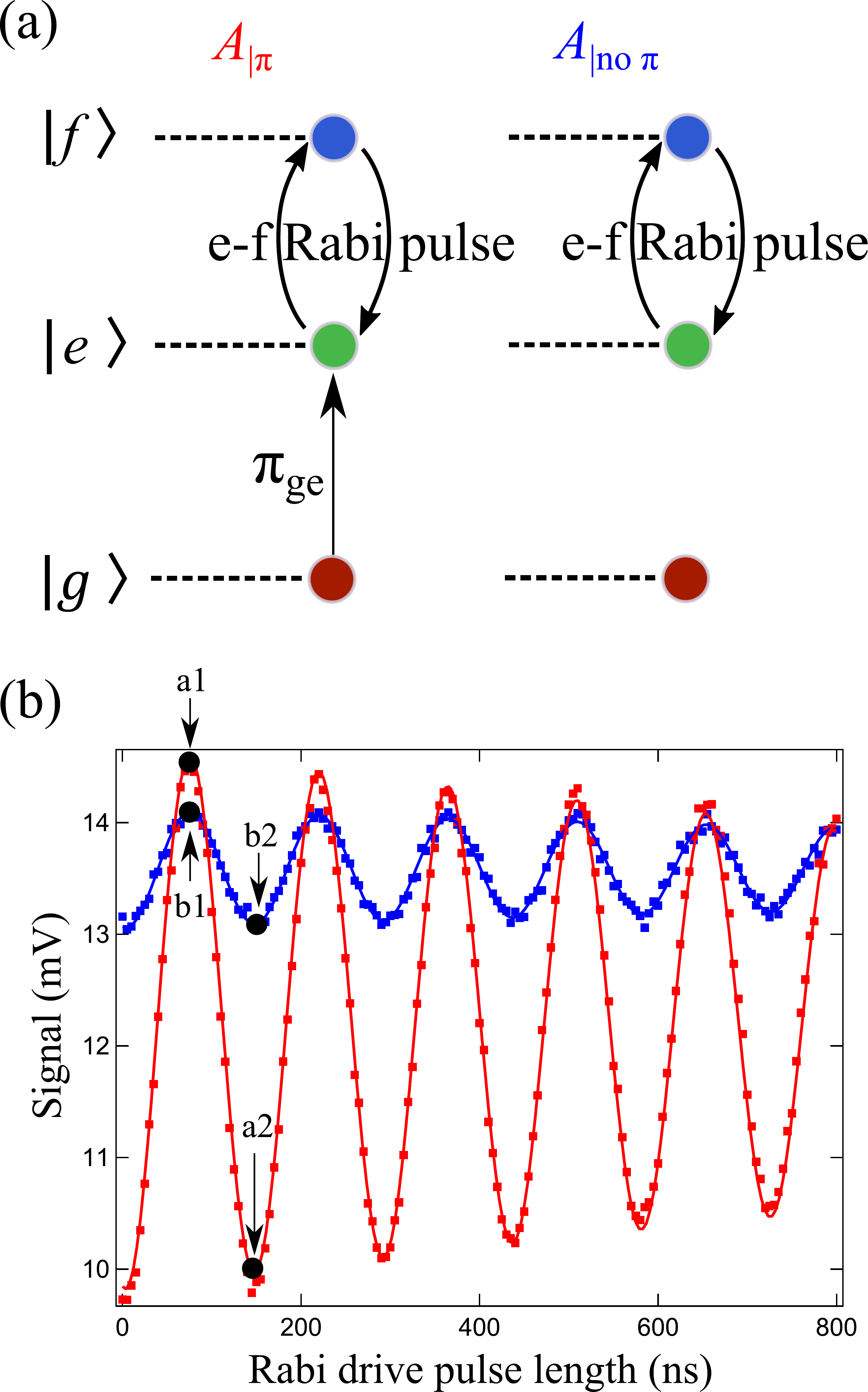}
        \caption{(a) Measurement scheme for the RPM method. 
        (b) Measurement results of scheme in (a).
        }
        \label{fig:RPM_theory}
\end{figure}

Here, we describe the Rabi population measurement (RPM). RPM is a method of measuring occupation by the protocol shown in Fig.~\ref{fig:RPM_theory}(a). The protocol assumes zero occupancies above the $\ket{f}$ state. The left panel in Fig.~\ref{fig:RPM_theory}(a) shows the measurement of the amplitude $A_{|\pi}$ of rabbinic oscillations due to the occupation difference between the $\ket{g}$ state and the $\ket{e}$ state. The panel on the right shows the measurement of the amplitude $A_{|{\rm no}\,\pi}$ of the rabi oscillation due to the occupation difference between the $\ket{e}$ state and the $\ket{f}$ state. The results of each measurement are shown in Fig.~\ref{fig:RPM_theory}(b). The amplitudes of the rabbinic oscillations are $A_{|\pi}=a1-a2$ and $A_{|{\rm no}\,\pi}=b1-b2$, and in the actual measurement, the occupation of the $\ket{e}$ state is calculated by solving Eq.~\eqref{RPM_theory} using measurement result at points a1, a2, b1, and b2.
\begin{equation}
\label{RPM_theory}
\mathcal{P}_{\ket{e}} = \frac{b1-b2}{(b1-b2)+(a1-a2)} \,.
\end{equation}

\subsection{Experimental setup}
\begin{figure*}[th!]
\centering
\includegraphics[width=120mm]{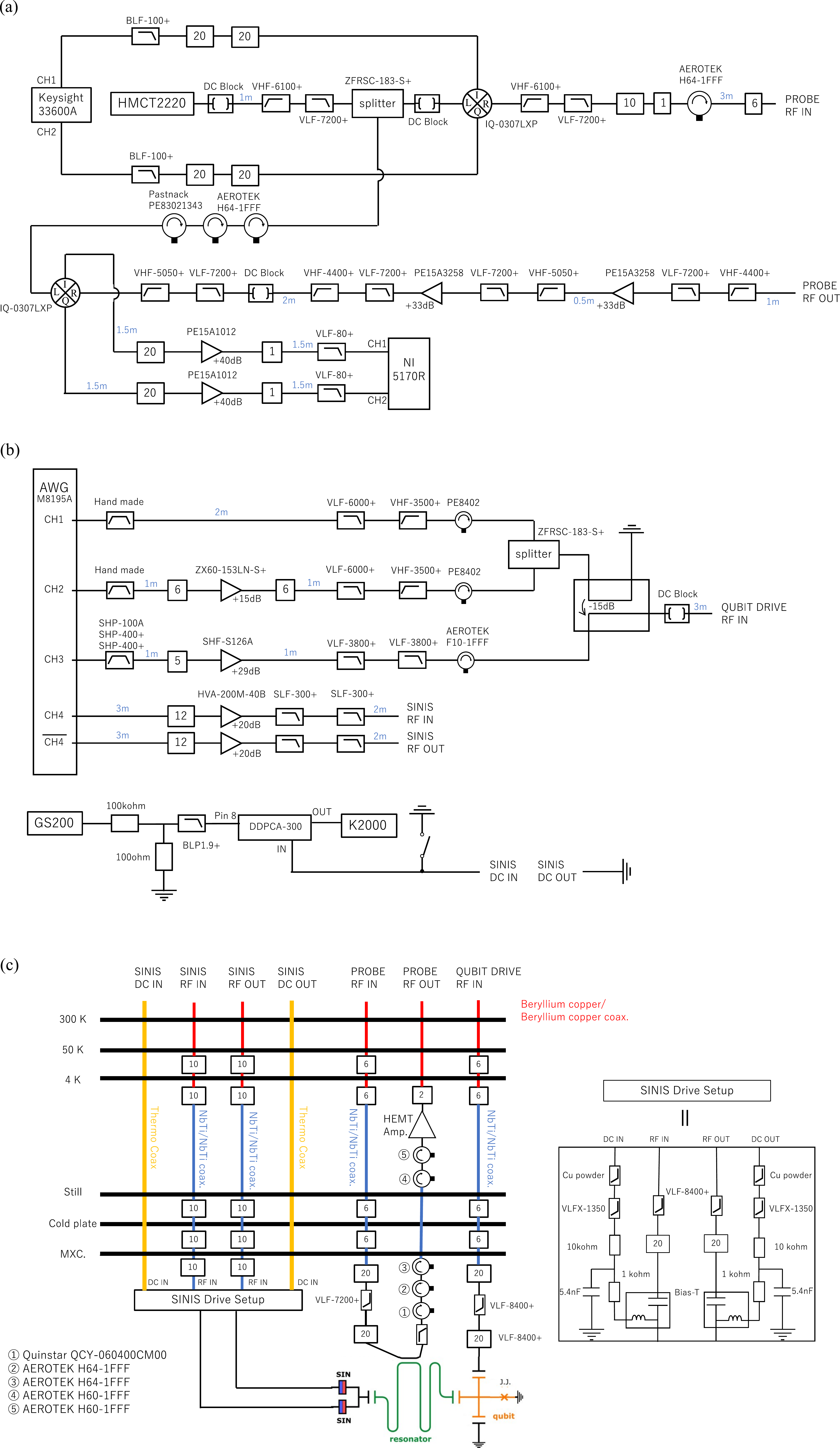}
        \caption{(a) Room temperature electronics setup for the readout line.
        (b) Room temperature electronics setup of drive pulse line of the qubit and bias pulse line and DC measurement line of QCR. 
        (c) Setup of dilution refrigerator.}
        \label{fig:setup}
\end{figure*}

The setup of the room temperature electronics and the internal wiring of the dilution refrigerator in this experiment are illustrated in Fig.~\ref{fig:setup}. The temperature of the dilution refrigerator during the measurement was approximately 12 mK.

\subsection{Master equation}
The following master equation was used for the numerical calculation of result in Fig.~\ref{fig:result}(a). This is taken from Ref.~\cite{doi:10.1063/5.0057894}. In the simulation of this experiment, the change in the relaxation rate of the resonator due to the QCR is calculated using Eq.~\eqref{resonator_transition} and the value is used as the damping term of the resonator to solve the master equation.

\begin{eqnarray}
\label{mastereq}
\dot{\rho}&=&-i/\hbar[\hat{H},\rho]\nonumber\\
&+&(\kappa_\mathrm{r}(1+N_\mathrm{tr})+\delta\gamma_\mathrm{QCR}(1+N_\mathrm{T}))\mathcal{D}[\hat{a}]\rho\nonumber\\
&+&(\kappa_\mathrm{r}N_\mathrm{tr}+\delta\gamma_\mathrm{QCR}N_\mathrm{T})\mathcal{D}[\hat{a}^{\dag}]\rho\nonumber\\
&+&(\gamma_\mathrm{Tge}(1+N_\mathrm{Tq})+\gamma_\mathrm{ge}(1+n_\mathrm{th}))\mathcal{D}[\ket{g}\bra{e}]\rho\nonumber\\
&+&(\gamma_\mathrm{Tge}N_\mathrm{Tq}+\gamma_\mathrm{ge}n_\mathrm{th})\mathcal{D}[\ket{e}\bra{g}]\rho\\
&+&(\gamma_\mathrm{Tef}(1+N_\mathrm{Tq})+\gamma_\mathrm{ef}(1+n_\mathrm{th}))\mathcal{D}[\ket{e}\bra{f}]\rho\nonumber\\
&+&(\gamma_\mathrm{Tef}N_\mathrm{Tq}+\gamma_\mathrm{ef}n_\mathrm{th})\mathcal{D}[\ket{f}\bra{e}]\rho\nonumber\\
&+&\gamma_\mathrm{\phi ge}\mathcal{D}[\ket{e}\bra{e}-\ket{g}\bra{g}]\rho\nonumber\\
&+&\gamma_\mathrm{\phi ef}\mathcal{D}[\ket{f}\bra{f}-\ket{e}\bra{e}]\rho\nonumber ¥,,
\end{eqnarray}
where $ \mathcal {D} [ \hat {A}] \rho = \hat {A} \rho \hat {A} ^ { \dag} - \{ \hat {A} ^ { \dag} \hat {A}, \rho\}/2 $. The first and second terms are damping terms due to resonator relaxation, the third and fourth terms are damping terms between the $\ket{g}$ and $\ket{e}$ states of the qubit, the fifth and sixth terms are damping terms between the $\ket{e}$ and $\ket{f}$ states, the seventh term is the phase relaxation term between the $\ket{g}$ and $\ket{e}$ states, and the eighth term is the phase relaxation term between the $\ket{e}$ and $\ket{f}$ states. Each parameter is also in accordance with Ref.~\cite{doi:10.1063/5.0057894}.


\begin{thebibliography}{99}
\bibitem{nakamura_coherent_1999} Y. Nakamura, Yu A. Pashkin, and J. S. Tsai, “Coherent control of macroscopic quantum states in a single-cooper-pair box,” Nature 398, 786–788 (1999).
\bibitem{Arute2019Quantum} Frank Arute, Kunal Arya, Ryan Babbush, Dave Bacon,
Joseph C Bardin, Rami Barends, Rupak Biswas, Sergio
Boixo, Fernando G S L Brandao, David A Buell, Brian
Burkett, Yu Chen, Zijun Chen, Ben Chiaro, Roberto
Collins, William Courtney, Andrew Dunsworth, Ed-
ward Farhi, Brooks Foxen, Austin Fowler, Craig Gidney,
Marissa Giustina, Rob Graff, Keith Guerin, Steve Habeg-
ger, Matthew P Harrigan, Michael J Hartmann, Alan
Ho, Markus Hoffmann, Trent Huang, Travis S Humble,
Sergei V Isakov, Evan Jeffrey, Zhang Jiang, Dvir Kafri,
Kostyantyn Kechedzhi, Julian Kelly, Paul V Klimov,
Sergey Knysh, Alexander Korotkov, Fedor Kostritsa,
David Landhuis, Mike Lindmark, Erik Lucero, Dmitry
Lyakh, Salvatore Mandr`a, Jarrod R McClean, Matthew
McEwen, Anthony Megrant, Xiao Mi, Kristel Michielsen,
Masoud Mohseni, Josh Mutus, Ofer Naaman, Matthew
Neeley, Charles Neill, Murphy Yuezhen Niu, Eric Os-
tby, Andre Petukhov, John C Platt, Chris Quintana,
Eleanor G Rieffel, Pedram Roushan, Nicholas C Ru-
bin, Daniel Sank, Kevin J Satzinger, Vadim Smelyanskiy,
Kevin J Sung, Matthew D Trevithick, Amit Vainsencher,
Benjamin Villalonga, Theodore White, Z Jamie Yao,
Ping Yeh, Adam Zalcman, Hartmut Neven, and
John M Martinis, “Quantum supremacy using a pro-
grammable superconducting processor,” Nature 574,
505–510 (2019).
\bibitem{GQA2021Exponential}Google Quantum AI, “Exponential suppression of bit or
phase errors with cyclic error correction,” Nature 595,
383–387 (2021).
\bibitem{PhysRevLett.127.180501}Yulin Wu, Wan-Su Bao, Sirui Cao, Fusheng Chen, Ming-
Cheng Chen, Xiawei Chen, Tung-Hsun Chung, Hui Deng,
Yajie Du, Daojin Fan, Ming Gong, Cheng Guo, Chu
Guo, Shaojun Guo, Lianchen Han, Linyin Hong, He-
Liang Huang, Yong-Heng Huo, Liping Li, Na Li, Shaowei
Li, Yuan Li, Futian Liang, Chun Lin, Jin Lin, Hao-
ran Qian, Dan Qiao, Hao Rong, Hong Su, Lihua Sun,
Liangyuan Wang, Shiyu Wang, Dachao Wu, Yu Xu, Kai
Yan, Weifeng Yang, Yang Yang, Yangsen Ye, Jianghan
Yin, Chong Ying, Jiale Yu, Chen Zha, Cha Zhang, Haibin
Zhang, Kaili Zhang, Yiming Zhang, Han Zhao, Youwei
Zhao, Liang Zhou, Qingling Zhu, Chao-Yang Lu, Cheng-
Zhi Peng, Xiaobo Zhu, and Jian-Wei Pan, “Strong
quantum computational advantage using a superconduct-
ing quantum processor,” Phys. Rev. Lett. 127, 180501
(2021).
\bibitem{divincenzo2000physical} David P DiVincenzo, “The physical implementation
of quantum computation,” Fortschritte der Physik:
Progress of Physics 48, 771–783 (2000).
\bibitem{Kwon2021Gate}Sangil Kwon, Akiyoshi Tomonaga, Gopika Lakshmi Bhai,
Simon J Devitt, and Jaw-Shen Tsai, “Gate-based su-
perconducting quantum computing,” Journal of applied
physics 129, 041102 (2021).
\bibitem{PhysRevLett.110.120501}K. Geerlings, Z. Leghtas, I. M. Pop, S. Shankar, L. Frunzio,
R. J. Schoelkopf, M. Mirrahimi, and M. H. Devoret,
“Demonstrating a driven reset protocol for a superconducting
qubit,” Phys. Rev. Lett. 110, 120501 (2013).
\bibitem{PhysRevApplied.10.044030}D.J. Egger, M. Werninghaus, M. Ganzhorn, G. Salis,
A. Fuhrer, P. M\"uller, and S. Filipp, “Pulsed reset pro-
tocol for fixed-frequency superconducting qubits,” Phys.
Rev. Applied 10, 044030 (2018).
\bibitem{reed2010fast}Matthew D Reed, Blake R Johnson, Andrew A Houck,
Leonardo DiCarlo, Jerry M Chow, David I Schuster,
Luigi Frunzio, and Robert J Schoelkopf, “Fast reset and
suppressing spontaneous emission of a superconducting
qubit,” Applied Physics Letters 96, 203110 (2010).
\bibitem{PhysRevB.101.235422} Hao Hsu, Matti Silveri, Andr\'as Gunyh\'o, Jan Goetz, Gi-
anluigi Catelani, and Mikko M\"ott\"onen, “Tunable re-
frigerator for nonlinear quantum electric circuits,” Phys.
Rev. B 101, 235422 (2020).
\bibitem{PhysRevApplied.17.044016} Y. Sunada, S. Kono, J. Ilves, S. Tamate, T. Sugiyama,
Y. Tabuchi, and Y. Nakamura, “Fast readout and reset
of a superconducting qubit coupled to a resonator with
an intrinsic purcell filter,” Phys. Rev. Applied 17, 044016
(2022).
\bibitem{tan2017quantum} Kuan Yen Tan, Matti Partanen, Russell E Lake, Joonas
Govenius, Shumpei Masuda, and Mikko M\"ott\"onen,
“Quantum-circuit refrigerator,” Nature communications
8, 1–8 (2017).
\bibitem{masuda2018observation} Shumpei Masuda, Kuan Y Tan, Matti Partanen, Rus-
sell E Lake, Joonas Govenius, Matti Silveri, Her-
mann Grabert, and Mikko M\"ott\"onen, “Observation
of microwave absorption and emission from incoherent
electron tunneling through a normal-metal–insulator–
superconductor junction,” Scientific reports 8, 1–8
(2018).
\bibitem{pekola2010environment}Jukka P Pekola, VF Maisi, Sergey Kafanov, Nikolai
Chekurov, A Kemppinen, Yu A Pashkin, O-P Saira,
M M\"ott\"onen, and JS Tsai, “Environment-assisted tun-
neling as an origin of the dynes density of states,” Phys-
ical Review Letters 105, 026803 (2010).
\bibitem{ingold1992charge}Gert-Ludwig Ingold and Yu V Nazarov, “Charge
tunneling rates in ultrasmall junctions,” in
Single charge tunneling (Springer, 1992) pp. 21–107.
\bibitem{sevriuk2019fast} VA Sevriuk, Kuan Yen Tan, Eric Hyypp\"a, Matti Sil-
veri, Matti Partanen, M\'at\'e Jenei, Shumpei Masuda, Jan
Goetz, Visa Vesterinen, Leif Gr\"onberg, et al., “Fast con-
trol of dissipation in a superconducting resonator,” Ap-
plied Physics Letters 115, 082601 (2019).
\bibitem{PhysRevLett.121.060502} P. Magnard, P. Kurpiers, B. Royer, T. Walter, J.-C.
Besse, S. Gasparinetti, M. Pechal, J. Heinsoo, S. Storz,
A. Blais, and A. Wallraff, “Fast and unconditional all-
microwave reset of a superconducting qubit,” Phys. Rev.
Lett. 121, 060502 (2018).
\bibitem{PhysRevB.100.134505} Matti Partanen, Jan Goetz, Kuan Yen Tan, Kassius
Kohvakka, Vasilii Sevriuk, Russell E. Lake, Roope
Kokkoniemi, Joni Ikonen, Dibyendu Hazra, Akseli
M\"akinen, Eric Hyypp\"a, Leif Gr\"onberg, Visa Vesterinen,
Matti Silveri, and Mikko M\"ott\"onen, “Exceptional points
in tunable superconducting resonators,” Phys. Rev. B
100, 134505 (2019).
\bibitem{sevriuk_initial_2022}V. A. Sevriuk, W. Liu, J. Rönkkö, H. Hsu, F. Marxer,
T. F. Mörstedt, M. Partanen, J. Räbinä, M. Venkatesh,
J. Hotari, L. Grönberg, J. Heinsoo, T. Li, J. Tuorila,
K. W. Chan, J. Hassel, K. Y. Tan, and M. Möttönen,
“Initial experimental results on a superconducting-qubit
reset based on photon-assisted quasiparticle tunneling,”
Applied Physics Letters 121, 234002 (2022), publisher:
American Institute of Physics.
\bibitem{doi:10.1063/5.0057894} T. Yoshioka and J. S. Tsai, “Fast unconditional initial-
ization for superconducting qubit and resonator using
quantum-circuit refrigerator,” Applied Physics Letters
119, 124003 (2021), https://doi.org/10.1063/5.0057894.
\bibitem{article} Matti Silveri, Hermann Grabert, Shumpei Masuda, Kuan
Tan, and Mikko M\"ott\"onen, “Theory of quantum-
circuit refrigeration by photon-assisted electron tun-
neling,” Physical Review B 96 (2017), 10.1103/Phys-
RevB.96.094524.
\bibitem{PhysRevLett.114.240501} X. Y. Jin, A. Kamal, A. P. Sears, T. Gudmundsen,
D. Hover, J. Miloshi, R. Slattery, F. Yan, J. Yoder, T. P.
Orlando, S. Gustavsson, and W. D. Oliver, “Thermal
and residual excited-state population in a 3d transmon
qubit,” Phys. Rev. Lett. 114, 240501 (2015).
\bibitem{silveri2019broadband} Matti Silveri, Shumpei Masuda, Vasilii Sevriuk, Kuan Y
Tan, M\'at\'e Jenei, Eric Hyypp\"a, Fabian Hassler, Matti
Partanen, Jan Goetz, Russell E Lake, et al., “Broadband
lamb shift in an engineered quantum system,” Nature
Physics 15, 533–537 (2019).
\bibitem{PhysRevResearch.3.033126} Arto Viitanen, Matti Silveri, M\'at\'e Jenei, Vasilii Sevriuk,
Kuan Y. Tan, Matti Partanen, Jan Goetz, Leif
Gr\"onberg, Vasilii Vadimov, Valtteri Lahtinen, and
Mikko M\"ott\"onen, “Photon-number-dependent effective
lamb shift,” Phys. Rev. Research 3, 033126 (2021).
\bibitem{jeffrey2014fast}Evan Jeffrey, Daniel Sank, J. Y. Mutus, T. C. White, J. Kelly, R. Barends, Y. Chen, Z. Chen, B. Chiaro, A. Dunsworth, A. Megrant, P. J. J. O’Malley, C. Neill, P. Roushan, A. Vainsencher, J. Wenner, A. N. Cleland, and John M. Martinis, “Fast accurate state measurement with super-
conducting qubits,” Physical review letters 112, 190504
(2014).
\bibitem{muhonen2012micrometre}Juha T Muhonen, Matthias Meschke, and Jukka P
Pekola, “Micrometre-scale refrigerators,” Reports on
Progress in Physics 75, 046501 (2012).
\end{thebibliography}
\end{document}